\documentstyle[12pt]{article}
\newcommand{\beq}{\begin{equation}}
\newcommand{\eeq}{\end{equation}}
\newcommand{\bea}{\begin{eqnarray}}
\newcommand{\eea}{\end{eqnarray}}

\begin{document}
\setcounter{page}{0}
\topmargin 0pt
\oddsidemargin 5mm
\renewcommand{\thefootnote}{\fnsymbol{footnote}}
\newpage
\setcounter{page}{0}
\begin{titlepage}
\begin{flushright}
QMW 94-6
\end{flushright}
\begin{flushright}
hep-th/9403097
\end{flushright}
\vspace{0.5cm}
\begin{center}
{\large {\bf The Gauged WZNW Model Perturbed By The Sigma Model Term}} \\
\vspace{1.8cm}
\vspace{0.5cm}
{\large Oleg A. Soloviev
\footnote{e-mail: soloviev@V1.PH.QMW.ac.uk}\footnote{Work supported by
S.E.R.C.}}\\
\vspace{0.5cm}
{\em Physics Department, Queen Mary and Westfield College, \\
Mile End Road, London E1 4NS, United Kingdom}\\
\vspace{0.5cm}
\renewcommand{\thefootnote}{\arabic{footnote}}
\setcounter{footnote}{0}
\begin{abstract}
{We discuss special perturbations of the gauged level $k$ WZNW model inspired
by the $\sigma$-model perturbation of the nonunitary WZNW model. In the large
$k$ limit there is a second conformal point in the vicinity of the ultarviolet
fixed point. At the second critical point the conformal model has a rational
Virasoro central charge which no longer corresponds to a coset construction. In
spite of this fact the perturbative conformal model appears to be a unitary
system as long as the underlying coset construction at the ultraviolet critical
point is unitary.}
\end{abstract}
\vspace{0.5cm}
\centerline{March 1993}
 \end{center}
\end{titlepage}
\newpage
\section{Introduction}

Coset constructions \cite{Bardakci_Halpern} provide a fairly
elegant description to a wide class of irreducible conformal representations.
In contrast with many known algebraic conformal systems, coset constructions
possess a proper Lagrangian formulation in terms of gauged
Wess-Zumino-Novikov-Witten (GWZNW) models \cite{Bardakci_Rabinovici}. The
significance of this can be appreciated, because many of GWZNW models admit
quite an interesting space-time interpretations giving rise to exact string
solutions \cite{Bardakci_Crescimanno}-\cite{Tseytlin}. This aspect of coset
constructions is nowadays under extensive consideration.

Of no less importance is another aspect related to the
Lagrangian formulation of coset constructions. Namely, this is the issue of
understanding cosets as critical points of larger (massive) systems.
This problem is in
the intimate connection to possible perturbations of GWZNW models by relevant
operators. Answering these questions might turn out to be helpful in
understanding the configuration space of string solutions which is believed to
form the coordinate space of string field theory.

In the present paper we are going to explore some special deformations of GWZNW
models inspired by the $\sigma$-model perturbation of nonunitary conformal
WZNW models \cite{Soloviev-1},\cite{Soloviev-2}. It was shown in
\cite{Soloviev-1},\cite{Soloviev-2} that the conformal WZNW model with large
negative level $k$ allows a relevant perturbation by the $\sigma$-model term.
The perturbed theory flows to the unitary conformal WZNW model with positive
level $|k|$. So that the two conformal points discovered by Witten
\cite{Witten-2} turn out to be the ultraviolet and infrared fixed points of the
one renormalization group flow.

In the case of the GWZNW model, the conformal nonunitary WZNW model emerges
within the path integral quantization of the gauge invariant system
\cite{Bardakci_Rabinovici}. Therefore, there is an opportunity to apply the
discovered perturbation of nonunitary conformal WZNW theories to unitary
systems. It is quite plausible that this way might lead us to some unitary
rational conformal models which are distinct from coset constructions.

\section{The $\sigma$-model perturbation of the GWZNW}

The crucial observation we are going to make use of is that at the classical
level the GWZNW model can be described as a combination of usual conformal
WZNW models
\begin{equation}
S_{GWZNW}=S_{WZNW}(hg\tilde h;k)~+~S_{WZNW}(h\tilde h;-k),\end{equation}
where $S_{WZNW}(f;l)$ is the level $l$ conformal WZNW model defined as follows
\begin{equation}
S_{WZNW}(f,l)=-{l\over 4\pi}\int
[\mbox{Tr}|f^{-1}\mbox{d}f|^2~+~{i\over3}\mbox{d}^{-1}\mbox{Tr}(f^{-1}
\mbox{d}f)^3].\end{equation}
Here the matrix field $f$ is either $g$ taking its values on the Lie group $G$
or $h,~\tilde h$ taking values on the subgroup $H$ of $G$. Respectively the
level $l$ is either $k$ or $-k$.

The action of the GWZNW model in terms of the group element $g$ and the vector
nonabelian fields $A_z$ and $\bar A_{\bar z}$ is obtained from eq. (2.1) upon
the following definition
\begin{equation}
A_z=h^{-1}\partial h,~~~~~~~~\bar A_{\bar z}=\bar\partial\tilde h\tilde
h^{-1}.\end{equation}
Correspondingly the gauge symmetry in the variables $g,~h,~\tilde h$ is given
by
\begin{equation}
g\to\Omega g\Omega^{-1},~~~~~h\to h\Omega^{-1},~~~~~\tilde h\to\Omega\tilde
h,\end{equation}
where $\Omega$ is the parameter of the gauge group $H$. Obviously the product
$h\tilde h$ is gauge invariant.
As a matter of fact, in the GWZNW model presented in the
variables $g,~h,~\tilde h$, the following operator
\begin{equation}
O=\mbox{const}:\mbox{Tr}[\partial(h\tilde h)\bar\partial(h\tilde h)^{-1}]:
\end{equation}
has to be a gauge invariant operator, although the very action of the fields
$h,~\tilde h$ becomes changed.

At the quantum level the action of the GWZNW in the variables
$g,~h,~\tilde h$ is given by \cite{Bardakci_Rabinovici}
\begin{equation}
S_{QGWZNW}=S_{WZNW}(hg\tilde h)~+~
S_{WZNW}(h\tilde h,-k-2c_V(H))~+~S_{Gh}(b,c,\bar b,\bar c).\end{equation}
Compared to the classical action in eq. (2.1), in the quantum action the second
WZNW model of the product $h\tilde h$ has the level shifted by two times the
eigenvalue of the quadratic Casimir operator in the adjoint representation of
the subalgebra ${\cal H}$. In addition, the QGWZNW model has the ghost-like
contribution
\begin{equation}
S_{Gh}=\mbox{Tr}\int d^2z(b\partial c~+~\bar b\partial\bar c).\end{equation}

The gauge symmetry (2.4) allows one to impose the following gauge condition
\begin{equation}
\tilde h=1,\end{equation}
which does not entail any additional propagating Faddeev-Popov ghosts. In this
gauge the operator $O$ in eq. (2.5) takes the form
\begin{equation}
O=\mbox{const}:\mbox{Tr}(\partial h\bar\partial h^{-1}):,\end{equation}
which is nothing but the $\sigma$-model term.

A convenient choice for the constant in eq. (2.9) is
\begin{equation}
\mbox{const}={(k+2c_V(H))^2\over 4}.\end{equation}
Then the operator $O$ can be written as a normal ordered product of the three
operators
\begin{equation}
O=:\phi^{a\bar a}\cdot J_a\cdot\bar J_{\bar a}:,\end{equation}
where
\begin{eqnarray}
\phi^{a\bar a}&=&\mbox{Tr}(h^{-1}t^aht^{\bar a}),\nonumber\\
J_a&=&{1\over 2}\eta_{ab}\mbox{Tr}[(k+2c_V(H))h^{-1}\partial ht^a],\\
\bar J_{\bar a}&=&{1\over 2}\eta_{\bar a\bar b}\mbox{Tr}[(k+2c_V(H))\bar
\partial hh^{-1}t^{\bar a}].\nonumber\end{eqnarray}
Here $t^a,~t^{\bar a}$ are the generators of two copies of the Lie algebra
${\cal H}$.

The important point to be made is that normal ordering in the product (2.11)
can be understood according to
\begin{equation}
O=\oint{dw\over2\pi}\oint{d\bar w\over2\pi}{J_a(w)\cdot\bar J_{\bar a}(\bar
w)\cdot\phi^{a\bar a}(z,\bar z)\over |z-w|^2},\end{equation}
where in the numerator the product is thought of as an operator product
expansion (OPE) with respect to the conformal WZNW model
$S_{WZNW}(h,-k-2c_V(H))$. Furthermore, one can compute the conformal dimensions
of the operator $O$
\begin{equation}
\Delta=\bar\Delta=1-c_V(H)/(k+c_V(H)).\end{equation}

{}From the last formula it follows that in the large $k$ limit, the given
operator is to be classified as a quasimarginal relevant operator. Besides, the
operator $O$ preserves explicitly the global $H\times H$ symmetry of the
conformal theory $S_{WZNW}(h,-k-2c_V(H))$. Because of this fact, the operator
$O$ has to obey the following fusion rule
\begin{equation}
O\cdot O=[O]~+~[I],\end{equation}
where the square brackets denote the contributions of $O$ and $I$ and the
corresponding descendants of $O$ and $I$. Here $I$ is identity operator.

Thus, the given operator $O$ has the properties which are appropriate for
performing a renormalizable perturbation of the conformal WZNW model
$S_{WZNW}(h,-k-2c_V(H))$. Since the latter appears as an intrinsic part of the
GWZNW model, the perturbation by the operator $O$ results in the perturbation
of the GWZNW theory. It is significant to point out that $O$ does not belong
to the subspace of operators of the $G/H$ coset. Indeed, one can check that
this operator $O$ is not
annihilated by the corresponding BRST operator of the GWZNW model
\cite{Karabali_Schnitzer}\footnote{ Note that the BRST operator selecting the
states corresponding to the $G/H$ coset is defined by the constraints which are
nothing but conditions of independence of the partition function of the GWZNW
model on the gauge fields $A_z,~\bar A_{\bar z}$\cite{Karabali_Schnitzer}.
Whereas the gauge invariance requires a more weak restriction. Namely,
\begin{eqnarray}
\partial<\bar J^{tot}>~-~\bar\partial<J^{tot}>=0,\nonumber\end{eqnarray}
where
\begin{eqnarray}
<\bar J^{tot}>={\delta Z({\cal A}_z,\bar{\cal A}_{\bar z})\over\delta
{\cal A}_z}|_{{\cal
A}_z,\bar{\cal A}_{\bar z}=0}=0,\;\;\;\;\;\;
<J^{tot}>={\delta Z({\cal A}_z,\bar{\cal A}_{\bar z})\over\delta\bar
{\cal A}_{\bar z}}|_{{\cal A}_z,\bar{\cal A}_{\bar
z}=0}=0.\nonumber\end{eqnarray}
Here ${\cal A}_z,~\bar{\cal A}_{\bar z}$ are background gauge fields associated
with the quantum $A_z,~\bar A_{\bar z}$.}.
At the same time, this operator $O$ being an affine
descendant of the affine-Virasoro primary field $\phi^{a\bar a}$:
\begin{eqnarray}
O=J_{-1}\bar J_{-1}\phi,\nonumber\end{eqnarray}
has positive norm (with respect to the $SL(2,C)$ invariant vacuum)
\begin{eqnarray}
||O||^2=||J_{-1}\bar
J_{-1}|\phi>||^2=(k+2c_V(H))^2||\phi||^2>0.\nonumber\end{eqnarray}
Furthermore, the Virasoro central charge of the nonunitary WZNW model
$S_{WZNW}(h,-k-2c_V(H))$ is positive. Indeed,
\begin{eqnarray}
c_{WZNW}(-k-2c_V(H))={(k+2c_V(H))\dim H\over k+c_V(H)}>\dim H.
\nonumber\end{eqnarray}
Thus, the conclusion is: since the operator $O$ has the positive conformal
dimensions given by eq.
(2.14), this operator provides a unitary representation of the Virasoro
algebra.

{}From now on by the
perturbed GWZNW model we will understand the following theory
\begin{equation}
S_{PGWZNW}=S_{QGWZNW}~-~\epsilon\int d^2z~O(z,\bar z).\end{equation}

\section{Perturbed conformal point}

We go on to compute the renormalization beta function corresponding to the
coupling $\epsilon$. Away of criticality, where $\epsilon\ne0$, the beta
function is defined according to (see e.g. \cite{Cardy})
\begin{equation}
\beta=[2-(\Delta+\bar\Delta)]\epsilon~-~\pi C~\epsilon~+~{\cal O}(\epsilon^3),
\end{equation}
where $\Delta,~\bar\Delta$ are given by eq. (2.14). The constant $C$ is the
coefficient of the three point function
\begin{equation}
<O(z_1)O(z_2)O(z_3)>=C~\Pi^3_{i<j}{1\over|z_{ij}|^{\Delta+\bar\Delta}}
\end{equation}
when the two point functions are normalized to unity.

The coefficient $C$ in the large $k$ limit is given by \cite{Soloviev-2}
\begin{equation}
C=c_V(H)~+~{\cal O}(1/k).\end{equation}
With the given $C$ one can easily solve equation (3.17) to find the
perturbative fixed point
\begin{equation}
\epsilon_2={-2\over\pi k}.\end{equation}
Note that the value of the perturbative conformal point in eq. (3.20)
can be corrected by higher orders in $1/k$. It is quite remarkable that one can
obtain the exact value of $\epsilon_2$. Indeed, the second conformal point
$\epsilon_2$ has to coincide with the second critical point of the WZNW model.
This is given as follows
\begin{equation}
\epsilon_2={-2\over\pi(k+2c_V(H))}.\end{equation}
Obviously in the large $k$ limit this solution goes to the perturbative
expression in eq. (3.20).

Once we know the perturbative critical point exactly, we can compute the
exact Virasoro central charge at this point $\epsilon_2$. We find
\begin{eqnarray}
c(\epsilon_2)&=&c(G/H)~+~c_{WZNW}(k+2c_V(H))~-~c_{WZNW}(-k-2c_V(H))\nonumber \\
& & \\
&=&
c(G/H)~-~{2(k+2c_V(H))c_V(H)\dim H\over k^2+4kc_V(H)+3c_V(H)^2}.\nonumber
\end{eqnarray}
Here $c_{WZNW}(l)$ denotes the Virasoro central charge of the level $l$ WZNW
model.
Thus we come to conclusion that the perturbative Virasoro central charge
$c(\epsilon_2)$ is less than $c(G/H)$.  According to Zamolodchikov's
$c$-theorem \cite{Zamolodchikov} the magnitude of the Virasoro central charge
at the infrared critical point, that is $\epsilon_2$, must be less than at
the ultraviolet critical point, $\epsilon=0$, if the flow between these points
is unitary. Since the flow in the case under consideration behaves in the full
agreement with the $c$-theorem, it is suggestive that the conformal system by
the perturbation of the GWZNW model gives rise to the unitary theory as long as
the GWZNW model describes a unitary coset. Note that it is not obvious from the
expression for
the central charge given by eq. (2.22) that the underlying conformal model
corresponds to another unitary coset construction. Thus it may turn out that
the BRST projection of GWZNW models is not only way of deriving unitary
Virasoro representations.

\section{Conclusion}

We have considered the perturbation of the $G/H$ GWZNW model by the gauge
invariant operator
coinciding with the $\sigma$-model term of the WZNW model on the subgroup $H$.
We exhibited that the perturbed GWZNW model arrives at the perturbative
(infrared)
critical point which has a rational Virasoro central charge. Surprisingly
this central charge does not appear to be a combination of
the Virasoro central charges of some unitary cosets, but we were not able
to prove it precisely. We found that
the flow agrees with the $c$-theorem for unitary systems. Based on it we
conjecture that the obtained
conformal system is unitary provided the $G/H$ coset projection of the GWZNW
model is unitary.

\par \noindent
{\em Acknowledgement}: I would like to thank J. M. Figueroa-O'Farrill,
C. M. Hull and I. Vaysburd
for useful
discussions. I would also like to thank the SERC for financial support.

\end{document}